\documentclass{article}
\usepackage{theorem}
\newcommand{\et}{\hspace*{1pt}\hfill\rule[-.2ex]{.8ex}{1.6ex}\vspace{.5ex}\pagebreak[3]}

\usepackage{graphicx}
\usepackage{amssymb}
\usepackage{graphicx}
\usepackage{cspsymb}
\usepackage{enumerate}
\usepackage{listings}
\usepackage{multirow}

\usepackage{url}
\urldef{\mailsa}\path|{alfred.hofmann, ursula.barth, ingrid.haas, frank.holzwarth,|
\urldef{\mailsb}\path|anna.kramer, leonie.kunz, christine.reiss, nicole.sator,|
\urldef{\mailsc}\path|erika.siebert-cole, peter.strasser, lncs}@springer.com|    

\newcommand{\treqrel}{\mathit{trace}}
\newcommand{\faeqrel}{\mathit{fail}}

\newcommand{\reveqrel}{\mathit{rev}}
\newcommand{\aceqrel}{\mathit{acc}}
\newcommand{\rteqrel}{\mathit{rt}}
\newcommand{\obs}{\mathrm{obs}}
\newcommand{\done}{\mathit{done}}
\newcommand{\stab}{\mathit{stab}}
\newcommand{\pref}{\mathrm{pref}}
\newcommand{\init}{\mathit{initials}}

\newcommand{\unstp}{\mathit{UNSTABLE}}
\newcommand{\stabp}{\mathit{STABLE}}
\newcommand{\prior}{\mathrm{\bf prioritise}}

\newcommand{\cfl}{\mathcal{C}_{\flmodel}}
\newcommand{\cm}{\mathcal{C}_{\mmodel}}
\newcommand{\PA}{P_{\mathcal{A}}}

\newcommand{\lft}{\mathrm{left}}
\newcommand{\rgt}{\mathrm{right}}

\newcommand{\Sym}{\mathit{Sym}}

\newtheorem{lemma}{Lemma}
\newtheorem{theorem}{Theorem}
\newtheorem{definition}{Definition}
\newtheorem{corollary}{Corollary}

\begin{document}

%\mainmatter  % start of an individual contribution

\title{Translating between models of concurrency}

%\titlerunning{Lecture Notes in Computer Science: Authors' Instructions}

	\author{David Mestel\thanks
	{Department of Computer Science, University of Luxembourg. Email: David.Mestel@gmail.com} \and
	\and A.W. Roscoe\thanks{%
	Department of Computer Science, University of Oxford.   Email: {Bill.Roscoe@cs.ox.ac.uk}}}
%
%\authorrunning{D. Mestel \and A.W. Roscoe}

%\address{Department of Computer Science, University of Oxford}%,\\
%\thanks[dmemail]{Email: \href{mailto:david.mestel@cs.ox.ac.uk}
%{\texttt{\normalshape david.mestel@cs.ox.ac.uk}}}
%\email{\{david.mestel, bill.roscoe\}@cs.ox.ac.uk}}
%\thanks[awremail]{Email: \href{mailto:bill.roscoe@cs.ox.ac.uk}
%{\texttt{\normalshape bill.roscoe@cs.ox.ac.uk}}}

	\maketitle

\begin{abstract}
Hoare's Communicating Sequential Processes
(CSP)~\cite{hoare1985communicating}
admits a rich universe of semantic models closely related to the van Glabbeek spectrum.  In this paper we study finite
observational models, of which at least six have been identified for CSP, namely
traces, stable failures, revivals, acceptances, refusal testing and finite linear
observations~\cite{roscoe2009revivals}.  We show how to use the
recently-introduced \emph{priority} operator (\cite{roscoe2010understanding},
ch.20) to transform refinement questions in these models into trace refinement
(language inclusion) tests.  Furthermore, we are able to generalise this
to any (rational) finite observational model.  As well as being of
theoretical interest, this is of practical significance since the
state-of-the-art refinement checking tool FDR4~\cite{fdr3} currently only
supports two such models.   In particular we study how it is possible to
check refinement in a discrete version of the Timed Failures model that
supports Timed CSP.
\end{abstract}
\section{Introduction}
In this paper we re-examine part of the Linear-Time spectrum that forms
part of the field of study of van Glabbeek in~\cite{vG1,vG2}, specifically the
part characterised by {\em finite linear} observations.

A number of different forms of process calculus have been developed for the
modeling of concurrent programs, including Hoare's Communicating Sequential
Processes (CSP)~\cite{hoare1985communicating}, Milner's Calculus of
Communicating Systems (CCS)~\cite{milner1986calculus}, and the
$\pi$-calculus~\cite{milner1992calculus}. Unlike the latter two, CSP's semantics
are traditionally given in behavioural semantic models coarser than
bisimulation, normally ones that depend on linear observations only.  Thus, while
the immediate range of options possible from a state can be observed, only one of them can
be followed in a linear observation and so branching behaviour is not recorded.

In this paper, we study finite\footnote{Models that use a mixture of finite and infinite linear behaviours
are also frequently used, the latter involving, {\em inter alia}, divergences and infinite traces.}
linear-time observational models for CSP; that
is, models where all observations considered can be determined in a finite time
by an experimenter who can see the visible events a process communicates and the sets of
events it can offer in any stable state.  While the experimenter can run the
process arbitrarily often, he or she can only record the results of individual
finite executions.  Thus each behaviour recorded can be deduced from a single
finite sequence of events and the visible events that link them, together with the sets of events accepted in stable
states during and immediately after this \emph{trace}. The representation
in the model is determined purely by the sent of these linear behaviours that it
is possible to observe of the process being examined.

At least six such models have been actively considered for CSP, but the state-of-the art
refinement checking tool, FDR4~\cite{fdr3,fdr3j}\footnote{See {\tt https://www.cs.ox.ac.uk/projects/fdr/}.  At the time of writing
there is no major academic paper as a source for FDR4 as opposed to its predecessor FDR3.}, currently only supports two, namely
\emph{traces} and \emph{ stable failures}\footnote{The word {\em stable} here
emphasises that the refusal components of failures are only recorded in stable (namely $\tau$-free) states.  This distinguishes it both from other models where
failures are recorded for other reasons: van Glabbeek (private correspondence) 
argues that there is in fact an {\em unstable} failures model (though anything but finite observation), citing~\cite{ufm} as evidence, and several versions have included failures on divergent traces or because of divergence-strictness.  We emphasise that all of the models considered in
this paper only observe acceptances and refusals in stable states.}.  FDR4 also supports the (divergence-strict) failures-divergences
model, which is not finite observational.

The question we address in this paper supposes that we have an automated proof
tool such as  FDR  that  answers questions  about  how
a process is represented in model A, and asks under what circumstances  it is
possible to answer  questions posed in model B, especially the core property of refinement.  

It seems intuitive that if model A records more details than model B, then
by looking carefully at how A codes the details recorded in B, 
the above ought to be possible.  We will later see some techniques for
achieving  this.   However it does not intuitively seem likely that
we can do the reverse.  Surprisingly, however, we find it can be done by
the use of process operators for which the coarser model B is not compositional.
Sometimes we can use such operators to transform observable features of behaviour that B does not see
into ones that it does.

The  operator we choose in  the world  of CSP is
 the relatively new {\em priority} operator. While simple
to define in operational semantics, this is only compositional over the finest
possible finite-linear-obseration model of CSP.  Priority is not part of
``standard'' CSP, but is implemented in the model checker FDR4 and greatly
extends the expressive power of the notation.  

We present first a construction which produces a context $\mathcal{C}$ such that
refinement questions in the well-known stable failures model correspond to trace refinement
questions under the application of $\mathcal{C}$.  We then generalise
this to show (Theorem \ref{thm:ratshift}) that a similar construction is possible not only for the
six models which have been studied, but also for any sensible finite
observational model (where `sensible' means
that the model can be recognised by a finite-memory computer, in a sense which
we shall make precise).  In fact we can seemingly handle any  equivalence determined in a compact way by  finitary  observations, even though not a congruence.

We first briefly describe the language of CSP.  We next give an informal
description of our construction for the stable failures model.  To prove the
result in full generality, we first give a formal definition of a finite
observational model, and of the notion of rationality.  We then describe our 
general construction.  
In a case study we consider the discrete version of the Timed Failures model
of Timed CSP, a closely related notation which already depends on priority
thanks to its need to enforce the principle of maximal progress.
For that we show not only how model shifting can obtain exactly what is needed
but also show how Timed Failures checking can be reduced to its relative
Refusal Testing.  Finally we discuss performance and optimisation issues.

The present paper is a revised and extended version of~\cite{msh}, with the
main additions being the study of Timed CSP and the model translation options
available there, plus a description of how to include CSP termination $\tick$.

\section{The CSP language}
We provide a brief outline of the language, largely taken from
\cite{roscoe2009revivals}; the reader is encouraged to consult
\cite{roscoe2010understanding} for a more comprehensive treatment.

Throughout, $\Sigma$ is taken to be a finite nonempty set of communications that
are visible and can only happen when the observing environment permits via
handshaken communication.  The actions of every process are taken from
$\Sigma\cup\{\tau\}$, where $\tau$ is the invisible internal action that cannot
be prevented by the environment.  We extend this to $\Sigma\cup\{\tau,\tick\}$ if we want the language to allow the successful termination process $\SKIP$
and sequential compositions as described below.  $\tick$ is different
from other events, because it is observable but not controllable: in that
sense it is between a regular $\Sigma$ event and $\tau$.  It only ever appears
at the end of traces and from a state which has refusal set $\Sigma$ and
acceptance set $\{\tick\}$, although that state is not stable in the
usual sense.  It thus complicates matters a little so the reader might
prefer to ignore it when first studying this paper.
We will later contemplate a second event with special semantics: $tock$
signifying the passage of time.

The constant processes of our core version of CSP are:
\begin{itemize}
\item $\STOP$ which does nothing---a representation of deadlock.
\item $\div$ which performs (only) an infinite sequence of internal $\tau$
actions---a representation of divergence or livelock.
\item $\CHAOS$ which can do anything except diverge, though this absence of
divergence is unimportant when studying finite behaviour models.
\item $\SKIP$ which terminates successfully.
\end{itemize}

The prefixing operator introduces communication: \begin{itemize}
\item $a\then P$ communicates the event $a$ before behaving like $P$.
\end{itemize}

There are two main forms of binary choice between a pair of processes:\begin{itemize}
\item $P\intchoice Q$ lets the process decide to behave like $P$ or like $Q$:
this is \emph{nondeterministic} or \emph{internal} choice.
\item $P\extchoice Q$ offers the environment the choice between the initial
$\Sigma$-events of $P$ and $Q$.  If the one selected is unambiguous then it
continues to behave like the one chosen; if it is an initial event of both then
the subsequent behaviour is nondeterministic.  The occurence of $\tau$ in one of
$P$ and $Q$ does \emph{not} resolve the choice (unlike CCS $+$).  This is
\emph{external} choice.
\end{itemize}

A further form of binary choice is the asymmetric
$P\rhd Q$, sometimes called {\em sliding} choice.  This offers any initial visible action of
$P$ from an unstable (in the combination) state and can (until such an action happens) perform
a $\tau$ action to $Q$.  It can be re-written in terms of prefix, external choice and hiding.  It represents a convenient shorthand way of creating
processes in which visible actions happen from an unstable state, so this is
not an operator one is likely to use much for building practical systems, rather
a tool for analysing how systems can behave.  As discussed in~\cite{roscoe2010understanding}, to give
a full treatment of CSP in any model finer than stable failures, it is necessary
to contemplate processes that have visible actions performed from unstable states.

We only have a single parallel operator in our core language since all the usual
ones of CSP can be defined in terms of it as discussed in Chapter 2 etc. of
\cite{roscoe2010understanding}.
\begin{itemize}
\item $P\parallel[X]Q$ runs $P$ and $Q$ in parallel, allowing each of them to
perform any action in $\Sigma\setminus X$ independently, whereas actions in $X$
must be synchronised between the two.
\end{itemize}

There are two operators that change the nature of a process's communications.
\begin{itemize}
\item $P\hide X$, for $X\subseteq\Sigma$, \emph{hides} $X$ by turning all $P$'s
$X$-actions into $\tau$s.
\item $P\rename{R}$ applies the \emph{renaming} relation
$R\subseteq\Sigma\times\Sigma$ to $P$: if $(a,b)\in R$ and $P$ can perform $a$,
then $P\rename{R}$ can perform $b$.  The domain of $R$ must include all visible
events used by $P$.  Renaming by the relation $\{(a,b)\}$ is 
denoted $\rensubs{a}{b}$.

\item Sequential composition $P;Q$ allows $P$ to run until it terminates successfully
($\tick$).  $P$'s $\tick$ is turned into $\tau$ and then $Q$ is started.
So if $P$ and $Q$ respectively have traces $s\cat\trace{\tick}$ and $t$,
then $P;Q$ has the trace $s\cat t$.
\end{itemize}

There is another operator that allows one process to follow another:
\begin{itemize}
	\item $P\:\Theta_{a:A} \:Q$ behaves like $P$ until an event in the set $A$ occurs, at
which point $P$ is shut down and $Q$ is started.  This is the \emph{throw}
operator, and it is important for establishing clean expressivity results.
\end{itemize}

The final CSP construct is \emph{recursion}: this can be single or mutual
(including mutual recursions over infinite parameter spaces), can be defined by
systems of equations or (in the case of single recursion) in line via the
notation $\rec{p}{P}$, for a term $P$ that may include the free process
identifier $p$.  Recursion can be interpreted operationally as having a
$\tau$-action corresponding to a single unwinding.  Denotationally, we regard
$P$ as a function on the space of denotations, and interpret $\rec{p}{P}$ as the
least (or sometimes provably unique) fixed point of this function.  

We also make use of the \emph{interleaving} operator $\interleave$, which allows
processes to perform actions independently and is equivalent to
$\parallel[\emptyset]$, and the process $\RUN_X$, which always offers every
element of the set $X$ and is defined by 
\[\RUN_X=\Extchoice_{x\in
X}x\then\RUN_X\]

This completes our list of operators other
than priority.  While others, for example $\triangle$ (interrupt) are sometimes
used, they are all expressible in terms of the above (see ch.9 of
\cite{roscoe2010understanding}).

\subsection{Priority}

The priority operator is introduced and discussed in detail in Chapter 20 of
\cite{roscoe2010understanding} as well as \cite{priexp}.  It allows us to specify an ordering on the
set of visible events $\Sigma$, and prevents lower-priority events from occuring
whenever a higher-priority event or $\tau$ is available.

The operator described in \cite{roscoe2010understanding} as implemented in
FDR4~\cite{fdr3} is parametrised by three arguments: a process $P$, a partial order
$\leq$ on the event set $\Sigma$, and a subset $X\subseteq\Sigma$ of events that
can occur when a $\tau$ is available.  We require that all elements of $X$ are
maximal with respect to $\leq$ and additionally require that if $a$  is any  event incomparable to $\tau$, then $a$ is also maximal.  Failing to respect these
principles means that the operator might undermine some basic principles of
CSP.
   Writing
$\init(P)\subseteq\Sigma\cup\{\tau\}$ for the set of events
that $P$ can immediately perform, and extending $\leq$ to a partial order on
$\Sigma\cup\{\tau\}$ by adding $y\leq\tau\forall y\in\Sigma\setminus X$, we define the operational semantics of
$\prior$ as follows:

\[\frac{P\trans[a] P'\land\forall b\neq a.a\leq
b\Rightarrow b\notin\init(P)}{\prior(P,\leq,X)\trans[a]\prior(P',\leq,X)}
\:(a\in\Sigma\cup\{\tau\}).\]

$\prior$ makes enormous contributions to the expressive power of CSP as explained in~\cite{priexp}, meaning that CSP+$\prior$ can be considered a universal
language for a much wider class of operational semantics than the {\em CSP-like}
class described in~\cite{cspexpr,roscoe2010understanding}.  

It should not therefore be surprising  that $\prior$ is not compositional over denotational finite observation models other than the
most precise model, as we will discuss below.
So we think of it as an optional addition to CSP rather than
an integral part of it; when we refer below to particular types of observation as giving
rise to valid
models for CSP, we will mean CSP without priority.

If we admit successful termination $\tick$, then it must have the same priority
as $\tau$.

\section{Example: the stable failures model}
We introduce our model shifting construction using the \emph{stable failures} model: we will produce
a context $\mathcal{C}$ such that for any processes $P,Q$, we have that $Q$
refines $P$ in the stable failures model if and only $\mathcal{C}[Q]$ refines
$\mathcal{C}[P]$ in the traces model.  

\subsection{The traces and failures models}
The \emph{traces} model $\tmodel$ is familiar from both process algebra and
 automata theory, and represents a process by
the set of (finite) strings of events it is able to accept.  Thus each process
is associated (for fixed alphabet $\Sigma$) to a subset of $\Sigma^*$ the set of
finite words over $\Sigma$ (plus words of the form $w\trace{\tick}$ if we
allow $SKIP$ and sequential composition).
The \emph{stable failures} model $\fmodel$ also records sets $X$ of events that the process is able
to stably refuse after a trace $s$ (that is, the process is able after trace $s$
to be in a state where no $\tau$ events are possible, and where the set of
initial events is disjoint from $X$).  Thus a process is associated to a subset of
$\Sigma^*\times (\mathcal{P}(\Sigma)\cup\{\spot\})$, where $\spot$ represents
the absence of a recorded refusal set.\footnote{This is equivalent to the
standard presentation in which a process is represented by a subset of
$\Sigma^*$ and one of $\Sigma^*\times\mathcal{P}(\Sigma)$: the trace component
is just $\{s:(s,\spot)\in\fmodel(P)\}$.}  We would add the symbol $\tick$ to this set when including termination.  Note that recording $\bullet$ does
not imply that there is no refusal to observe, simply that we have not observed
stability.  The observation of the refusal $\emptyset$ implies that the process
can be stable after the present trace, whereas observing $\bullet$ does not.

In any model $\mmodel$, we say that $Q$ \emph{$\mmodel$-refines} $P$, and write
$P\refinedby[M] Q$, if the set associated to $Q$
is a subset of that corresponding to $P$.

Because $\tick$ can be seen, but happens automatically, we need to distinguish
a process like $\SKIP$ which must terminate from one that can but may not like
$\STOP\intchoice\SKIP$.  After all if these are subsituted for $P$ in
$P;Q$ we get processes equivalent to $Q$ and $\STOP\intchoice Q$.  However
the state that accepts $\tick$ can be thought of as being able to refuse the
rest of the visible events $\Sigma$, since it can terminate all by itself.

\subsection{Model shifting for the stable failures model} We first
consider this without $\tick$. 
The construction is as
follows: 
\begin{lemma}For each finite alphabet $\Sigma$ there exists a context
$\mathcal{C}$ (over an expanded alphabet) such that for any processes $P$ and
$Q$ we have that $P\refinedby[F] Q$ if and only if
$\mathcal{C}[P]\refinedby[T]\mathcal{C}[Q]$.\end{lemma} 

\noindent{\bf Proof}\\
\textbf{Step 1:} We use priority to produce a process (over an expanded
alphabet) that can
communicate an event $x'$ if and only if the original process $P$ is able to
stably refuse $x$.  

This is done by expanding the alphabet $\Sigma$ to
$\Sigma\cup\Sigma'$ (where $\Sigma'$ contains a corresponding primed event $x'$ for
every event  $x\in\Sigma$), and prioritising with respect to the partial order which
prioritises each $x$ over the corresponding $x'$ and makes $\tau$ incomparable to
$x$ and greater than $x'$.   

We must also introduce an event $\stab$ to signify the observation of
 stability (i.e. no $\tau$ is possible in this state)
 without requiring
any refusals to be possible.  This is necessary in order to be able to record an empty refusal set.
The priority order $\leq_1$ is then the above (i.e. $x'<x$ for all $x\in\Sigma$) extended by making  $stab$ less
than only $\tau$ and independent of  all $x$ and $x'$.

We can now fire up these new events as follows:
\[\mathcal{C}_1[P] = \prior(P\interleave \RUN_{\Sigma'\cup\{\stab\}},\leq_1,\Sigma).\]
This process has a state $\xi'$ for each state $\xi$ of $P$, where $\xi'$ has the
same unprimed events (and corresponding transitions) as $\xi$.  Furthermore $\xi'$ can
communicate $x'$ just when $\xi$ is stable and can refuse $X$, and $\stab$ just
when $\xi$ is stable.

\textbf{Step 2:} We now recall that the definition of the stable failures model only
allows a refusal set to be recorded at the \emph{end} of a trace, and is not
interested in (so does not record) what happens after the refusal set.  

We gain this effect by using a regulator process to prevent a primed event (or
$\stab$) from being followed by an unprimed event.  Let 
\[\begin{array}{rl}
	\unstp = &\Extchoice_{x\in\Sigma} x\then \unstp \\
\extchoice &\Extchoice_{x\in\Sigma'\cup\{\stab\}} x \then \stabp \\
\stabp = &\Extchoice_{x\in\Sigma'\cup\{\stab\}} x \then\stabp,
\end{array}\]
and define $\mathcal{C}$ by 
\[\mathcal{C}[P] = \mathcal{C}_1[P] \parallel[\Sigma\cup\Sigma' \cup\{\stab\}]
\unstp.\]

A trace of $\mathcal{C}[P]$ consists of: firstly, a trace $s$ of $P$; followed by,
if $P$ can after $s$ be in a stable state, then for some such state $\sigma_0$ any string
formed from the events that can be refused in $\sigma_0$, together with $\stab$.
The lemma clearly follows. \et

It is clear that any such context must involve an operator that is
not compositional over traces, for otherwise we would have $P\trefinedby Q$
implies $\mathcal{C}[P]\trefinedby \mathcal{C}[Q]$, which is equivalent to
$P\frefinedby Q$, and this is not true for general $P$ and $Q$ (consider for
instance $P=a\then\STOP$, $Q=(a\then\STOP )\intchoice\STOP$).  It follows that
only contexts which like ours involve priority or some operator with similar
status can achieve this.

Adding $\tick$ to the model causes a few issues with the above.  For one thing
it creates a refusal (namely of everything except $\tick$) from what could be an unstable
state, namely a state that can perform $\tick$ and perhaps also a $\tau$.
And secondly we need to find an effective way of making processes show their
refusal of $\tick$, and their refusal of all events other than $\tick$, when
respectively appropriate.  One way of doing these things is to add to the
state space so that termination goes through multiple stages.  Create a new
event $term$ and consider $P;term\then\SKIP$.  This performs any behaviour
of $P$ except that all $\tick$s of $P$ become $\tau$s and lead to $term\then\SKIP$.  That of course is a stable state.   If we now (treating $term$ as
a member of $\Sigma$) apply $\mathcal C$ as defined above, this will
be able to perform $term'$ in any stable state that cannot terminate, and will perform every
$a'$ event other than $term'$ every time it reaches the state $term\then\SKIP$.
Thus if we define
\[\mathcal{C}^{\tick}(P) = \mathcal{C}(P;term\then\SKIP)\hide term\]
we get exactly the decorated traces we might have expected from the stable
failures representation of $P$ except that instead of having an event $\tick'$
we have $term'$.

%Note that efficiency can be improved by only allowing each refusal event to be
%communicated once and only in alphabetical order, but this is not required for
%correctness.

\section{Semantic models}
In order to generalise this construction to arbitrary finite observational
semantic models, we must give formal definitions not only of particular models
but of the very notion of a finite observational model.

\subsection{Finite observations}
We consider only models arising from \emph{finite linear observations}.
Intuitively, we postulate that we are able to observe the process performing a
finite number of visible actions, and that where the process was stable (unable
to perform a $\tau$) immediately before an action, we are able to observe the
\emph{acceptance set} of actions it was willing to perform. 

Note that there
cannot be two separate stable states before visible event $b$ without  another
visible event $c$ between them, even though it is possible to have many
visible events  between  stable states.  Thus it makes no sense to record two
separate refusals or acceptance sets between consecutive visible events.  
Similarly it does not make sense to record both an acceptance and a refusal, 
since observing an acceptance set means that recording a refusal conveys no
extra information: if acceptance $A$ is observed then no other is seen before
the next visible event, and observable refusals are exactly those disjoint from
$A$.

We are unable to finitely observe \emph{instability}: the most we are able to
record from an action in an unstable state is that we did not \emph{observe}
stability.  Thus in any context where we can observe stability we can also fail
to observe it by simply not looking.

We take models to be defined over finite alphabets $\Sigma$, and take an
arbitrary linear ordering on each finite $\Sigma$ to be \emph{alphabetical}.

The most precise finite observational model is that considering all finite
linear observations, and is denoted $\flmodel$:

\begin{definition}\label{def:flobs}The set of \emph{finite linear observations} over an
alphabet $\Sigma$ is
\[\flmodel_\Sigma:=\{\seq{A_0,a_1,A_1,\ldots,A_{n-1},a_n,A_n}:n\in \mathbb{N}, 
a_i\in \Sigma, A_i \subseteq \Sigma{\tt or } A_i=\spot\},\]
where the $a_i$ are interpreted as a sequence of communicated events, and the
$A_i$ denote stable acceptance sets, or in the case of $\spot$ failure to
observe stability.  Let the set of such observations corresponding to a process
$P$ be denoted $\flmodel_\Sigma(P)$.  This needs to be extended to encompass
final $\tick$s if we want to include termination.\end{definition}

(Sometimes we will drop the $\Sigma$ and just write $\flmodel(P)$).

More formally, $\flmodel(P)$ can be defined inductively; for instance
\[\flmodel(P\extchoice Q):=\left\{\seq{A\union B}\cat\alpha, \seq{A\union
B}\cat\beta : \seq{A}\cat\alpha\in \flmodel(P), \seq{B}\cat\beta \in
\flmodel(Q)\right\}\]
(where $X\cup\bullet := \bullet$ for any set $X$).  See Section 11.1.1 of
\cite{roscoe2010understanding} for further details.

Observe that $\flmodel$ has a natural partial order corresponding to extensions 
(where $\alpha\cat\langle\spot\rangle\cat\beta$ and $\alpha\cat\langle
A\rangle$
are both extended by $\alpha\cat\langle A\rangle\cat\beta$ for any set
$A$ and any $\alpha$ and $\beta$).  Note that for any process $P$ we have that $\flmodel(P)$ is downwards-closed
with respect to this partial order.

The definition of priority over  $\flmodel$ (accommodating final $\tick$s) is as follows. 
 $\prior(P,\leq,X)$ is,
with $\leq$ extended to the whole  of $\Sigma\cup\{\tau\}$ by making all elements not in $X$ incomparable to  all others
\[\begin{array}{c}
\{
\trace{A_0,b_1,A_1,\ldots,A_{n-1},b_n,A_n}\mid
\trace{Z_0,b_1,Z_1,\ldots,Z_{n-1},b_n,Z_n}\in P\}
\\
\cup
\\
\{
\trace{A_0,b_1,A_1,\ldots,A_{n-1},b_n,\bullet,\tick}\mid
\trace{Z_0,b_1,Z_1,\ldots,Z_{n-1},b_n,\bullet,\tick}\in P\}
\end{array}
\]
where for each $i$ one of the following holds:
\begin{itemize}
\item $b_i$ is maximal under $\leq$ and $A_i=\bullet$ (so there is no
condition on $Z_i$ except that it exists).
\item $a_i$ is not maximal under $\leq$ and $A_{i-1}=\bullet$ and $Z_i$ is
not $\bullet$ and neither does $Z_i$ contain any $c>b_i$.
\item Neither $A_i$ nor $Z_i$ is $\bullet$, and $A_i=\{a\in Z_i\mid \neg\exists b\in Z_i.b>a\}$,
\item and in each case where $A_{i-1}\neq\bullet$,  $a_i\in A_{i-1}$.
\end{itemize}

This is not possible for the other studied finite behaviour models of CSP:
the statement that it is for refusal testing $\rtmodel$ in~\cite{roscoe2010understanding} is not true,
though it is possible for some partical orders $\leq$ including those needed
for maximal progress in timed modelling of the sort we will see later.

\subsection{Finite observational models}
We consider precisely the models which are derivable from the observations of $\flmodel$, which
are well-defined in the sense that they are compositional over CSP syntax (other
than priority), and
which respect extension of the alphabet $\Sigma$.

\begin{definition}\label{def:premodel}A finite observational \emph{pre-model} $\mmodel$ consists for each
(finite) alphabet $\Sigma$ of a set of \emph{observations}, $\obs_\Sigma(\mmodel)$,
together with a relation
$\mmodel_\Sigma\subseteq\flmodel_\Sigma\times\obs_\Sigma(\mmodel)$.  The
representation of a process $P$ in $\mmodel_{\Sigma}$ is denoted
$\mmodel_\Sigma(P)$, and is given by 
\[\mmodel_\Sigma(P):=\mmodel_\Sigma(\flmodel_\Sigma(P))=\{y\in\obs_\Sigma(\mmodel):\exists
x\in\flmodel_\Sigma(P).(x,y)\in\mmodel_\Sigma\}.\]
For processes $P$ and $Q$ over alphabet $\Sigma$, if we have
$\mmodel_\Sigma(Q)\subseteq\mmodel_\Sigma(P)$ then we say $Q$
\emph{$\mmodel$-refines} $P$, and write
$P\refinedby[M]Q$.
\end{definition}
(As before we will sometimes drop the $\Sigma$).

Note that this definition is less general than if we had defined a pre-model to
be any
equivalence relation on $\mathcal{P}\left(\flmodel_\Sigma\right)$.  For
example, the equivalence relating sets of the same cardinality has no
corresponding pre-model.  Definition \ref{def:premodel} agrees with that sketched in
\cite{roscoe2010understanding}.

Without loss of generality, $\mmodel_\Sigma$ does not identify any elements of
$\obs_\Sigma(\mmodel)$; that is, we have
$\mmodel_\Sigma^{-1}(x)=\mmodel_\Sigma^{-1}(y)$ only if $x=y$ (otherwise
quotient by this equivalence relation).  Subject to this assumption,
$\mmodel_\Sigma$ induces a partial order on $\obs_\Sigma(\mmodel)$:

\begin{definition}The partial order \emph{induced by $\mmodel_\Sigma$ on
$\obs_\Sigma(\mmodel)$} is given by: $x\leq y$ if and only if for all
$b\in\mmodel_\Sigma^{-1}(y)$ there exists $a\in\mmodel_\Sigma^{-1}(x)$ with
$a\leq b$.\end{definition}

Observe that for any process $P$ it follows from this definition that
$\mmodel(P)$ is downwards-closed with respect to this partial order (since
$\flmodel(P)$ is downwards-closed).

\begin{definition}\label{def:comp}A pre-model $\mmodel$ is \emph{compositional} if for
all CSP operators $\bigoplus$, say of arity $k$, and for all processes
$P_1,\ldots,P_k$ and $Q_1,\ldots,Q_k$ such that $\mmodel(P_i)=\mmodel(Q_i)$ for
all $i$, we have \[\mmodel\left(\bigoplus(P_i)_{i=1\ldots k}\right) =
\mmodel\left(\bigoplus(Q_i)_{i=1\ldots k}\right).\]\end{definition}

This means that the operator defined on processes in $\obs(\mmodel)$ by taking the
pushforward of $\bigoplus$ along $\mmodel$ is well-defined: for any sets
$X_1,\ldots,X_k\subseteq\obs(\mmodel)$ which correspond to the images of CSP
processes, take processes $P_1,\ldots,P_k$ such that $X_i=\mmodel(P_i)$, and let 
\[\bigoplus(X_i)_{i=1\ldots k} = \mmodel\left(\bigoplus(P_i)_{i=1\ldots
k}\right).\]
Definition \ref{def:comp} says that the result of this does not depend on the
choice of the $P_i$.

Note that it is not necessary to require the equivalent of Definition
\ref{def:comp} for recursion in the
definition of a model, because of the following lemma which shows that least
fixed point recursion is automatically well-defined (and formalises some
arguments given in \cite{roscoe2010understanding}):

\begin{lemma}Let $\mmodel$ be a compositional pre-model.  Let $\mathcal{C}_1,\mathcal{C}_2$ be CSP contexts, such that for
any process $P$ we have $\mmodel(\mathcal{C}_1[P])=\mmodel(\mathcal{C}_2[P])$.  Let the
least fixed points of $\mathcal{C}_1$ and $\mathcal{C}_2$ (viewed as functions
on $\mathcal{P}(\flmodel)$ under the subset order) be $P_1$ and $P_2$
respectively.  Then $\mmodel(P_1)=\mmodel(P_2)$.\end{lemma}

\noindent{\bf Proof}\\
Using the fact that CSP contexts induce Scott-continuous functions on
$\mathcal{P}(\flmodel)$ (see \cite{hoare1985communicating}, Section 2.8.2), the
Kleene fixed point theorem gives that
$P_i=\bigcup_{n=0}^{\infty}\mathcal{C}_i^n(\bot)$.
Now any $x\in\mmodel(P_1)$ is in the union taken up to some finite $N$, and
since finite unions correspond to internal choice, and $\bot$ to the process
$\div$, we have that the unions up to $N$ of $\mathcal{C}_1$ and $\mathcal{C}_2$
agree under $\mmodel$ by compositionality.  Hence $x\in\mmodel(P_2)$, so
$\mmodel(P_1)\subseteq\mmodel(P_2)$.  Similarly
$\mmodel(P_2)\subseteq\mmodel(P_1)$.
%
%We treat the case of unary contexts; higher arities follow similarly operating
%on appropriate direct products.
%
%First recall that if $\mathcal{C}_1,\mathcal{C}_2$ represent CSP contexts, then
%they induce Scott-continuous functions on $\mathcal{P}(\flmodel)$ (See
%\cite{hoare1985communicating}, Section 2.8.2).  Then by the
%Kleene fixed point theorem, we have that 
%\[P_i=\Lor_{n=0}^{\infty}\mathcal{C}_i^n(\bot).\]
%
%Consider any $x\in\mmodel(P_1)$.  Then there exists some $N$ such that 
%\[x\in\mmodel\left(\Lor_{n=1}^N\mathcal{C}_1^n(\bot)\right).\]
%Observe that finite joins in $\flmodel$ correspond to the CSP operator of finite nondeterminsitic
%choice, and that $\bot$ corresponds to the process $\div$ (of which no finite
%observation can be made other than $\spot$), so that from the
%hypothesis of the lemma and trivial induction we have for any $n$ that
%$\mmodel(\mathcal{C}_1^n(\bot))=\mmodel(\mathcal{C}_2^n(\bot))$. Then by
%compositionality, we have 
%\[\mmodel\left(\Lor_{n=1}^N\mathcal{C}_1^n(\bot)\right) =
%\mmodel\left(\Lor_{n=1}^N\mathcal{C}_2^n(\bot)\right) \subseteq \mmodel(P_2),\]
%and so $x\in\mmodel(P_2)$.  Hence $\mmodel(P_1)\subseteq\mmodel(P_2)$, and
%similarly we have $\mmodel(P_2)\subseteq\mmodel(P_1)$, so
%$\mmodel(P_1)=\mmodel(P_2)$ as required.\qed
\et

\begin{definition}A pre-model $\mmodel$ is \emph{extensional} if for all
alphabets $\Sigma_1\subseteq\Sigma_2$ we have that
$\obs_{\Sigma_1}(\mmodel)\subseteq\obs_{\Sigma_2}(\mmodel)$, and
$\mmodel_{\Sigma_2}$ agrees with
$\mmodel_{\Sigma_1}$ on $\flmodel(\Sigma_1)\times\obs_{\Sigma_1}(\mmodel)$.\end{definition}

\begin{definition}\label{def:mod}A pre-model is a \emph{model} if it is compositional and
extensional.\end{definition}

In this setting, we now describe the five main finite observational models
coarser than $\flmodel$: traces, stable failures, revivals, acceptances and refusal
testing.

\subsubsection{The traces model}\quad
The coarsest model measures only the \emph{traces} of a process; that is, the sequences
of events it is able to accept.  This corresponds to the language of the process
viewed as a nondeterministic finite automaton (NFA).

\begin{definition}\label{def:trmod}The \emph{traces} model, $\tmodel$, is given by 
\[\obs_\Sigma(\tmodel) = \Sigma^*,\;\tmodel_\Sigma = \treqrel_\Sigma\]
where $\treqrel$ is the equivalence relation which relates the observation \\
$\seq{A_0,a_1,A_1,\ldots,a_n,A_n}$ to the string $a_1\ldots a_n$.\end{definition}

\subsubsection{Failures}\quad
The traces model gives us information about what a process is \emph{allowed} to
do, but it in some sense tells us nothing about what it is \emph{required} to
do.  In particular, the process $\STOP$ trace-refines any other process.  

In
order to specify liveness properties, we can incorporate some information about
the events the process is allowed to refuse, begining with the \emph{stable failures} model.
Intuitively, this captures traces $s$, together with the sets of events the
process is allowed to stably refuse after $s$.

\begin{definition}\label{def:famod}The \emph{stable failures} model, $\fmodel$, is given by
\[\obs_\Sigma(\fmodel) = \Sigma^*\times(\mathcal{P}(\Sigma)\cup\{\spot\}),\;\fmodel_\Sigma =
\faeqrel_\Sigma,\]
where $\faeqrel_\Sigma$ relates the observation $\seq{A_0,\ldots,a_n,A_n}$ to all 
pairs $(a_1\ldots a_n,X)$, for all $X\subseteq\Sigma\setminus A_n$ if
$A_n\neq\,\spot$, and for $X=\,\spot$ otherwise.\end{definition}

\subsubsection{Revivals}\quad
The next coarsest model, first introduced in \cite{roscoe2009revivals}, is the
\emph{revivals} model.  Intuitively this captures traces $s$, together with sets $X$
that can be stably refused after $s$, and events $a$ (if any) that can then be accepted.

\begin{definition}The \emph{revivals} model, $\rmodel$, is given by 
\[\obs_\Sigma(\rmodel) = \Sigma^*\times (\mathcal{P}(\Sigma)\cup\{\spot\})\times
(\Sigma\cup\{\spot\}),\;\rmodel_\Sigma = \reveqrel_\Sigma),\]
where $\reveqrel_\Sigma$ relates the
observation $\seq{A_0,a_1,\ldots,a_{n-1},A_{n-1},a_n,A_n}$ to \begin{enumerate}[(i)] \item the
triples $(a_1\ldots a_{n-1},X,a_n)$, for all $X\subseteq\Sigma\setminus
A_{n-1}$ if $A_{n-1}\neq\,\spot$ and for $X=\,\spot$ otherwise, and \item the
triples $(a_1\ldots a_n,X,\spot)$, for all $X\subseteq\Sigma\setminus A_n$ if
$A_n\neq\,\spot$ and for $X=\,\spot$ otherwise.\end{enumerate}\end{definition}

A finite linear observation is related to all triples consisting of: its initial
trace; a stable refusal that could have been observed, or $\spot$ if the
original observation did not observe stability; and
optionally (part (i) above) a single further event that can be accepted.

\subsubsection{Acceptances}\quad
All the models considered up to now refer only to sets of refusals, which in
particular are closed under subsets.  The next model, \emph{acceptances} (also
known as `ready sets'), refines the previous three and also 
considers the precise sets of events that can be stably accepted at the ends of
traces.

\begin{definition}The \emph{acceptances} model, $\amodel$, is given by 
\[\obs_\Sigma(\amodel) = \Sigma^*\times(\mathcal{P}(\Sigma)\cup\{\spot\}),\;
\amodel_\Sigma = \aceqrel_\Sigma,\]
where $\aceqrel_\Sigma$ relates the observation $\seq{A_0,a_1,\ldots,a_n,A_n}$ to
the pair $(a_1\ldots a_n,A_n)$.\end{definition}

It is convenient to note here that, just as we were able to use $a'$ as
a cipher for the refusal of $a$ when model shifting, we can introduce a second
	one $a''$ as a chipher for stable acceptance of $a$: it is performed (without changing the state) just when $a'$ is stably refused.  We will apply
	this idea and discuss it further below.

\subsubsection{Refusal testing}\quad
The final model we consider is that of \emph{refusal testing}, first introduced
in \cite{phillips1987refusal}.  This refines
$\fmodel$ and $\rmodel$ by considering an entire history of events and stable
refusal sets.  It is incomparable to $\amodel$, because it does not capture
precise acceptance sets.

\begin{definition}\label{def:rtmod}The \emph{refusal testing} model, $\rtmodel$, is given by
\[\obs_\Sigma(\rtmodel) =
\{\seq{X_0,a_1,X_1,\ldots,a_n,X_n}:n\in\mathbb{N},a_i\in\Sigma,
X_i\subseteq\Sigma{\tt or }X_i=\spot\}\]
\[\rtmodel_\Sigma = \rteqrel_\Sigma,\]
where $\rteqrel_\Sigma$ relates the observation $\seq{A_0,\ldots,a_n,A_n}$ to 
$\seq{X_0,\ldots,a_n,X_n}$, for all $X_i\subseteq\Sigma\setminus A_i$ if
$A_i\neq\,\spot$, and for $X_i=\,\spot$ otherwise.
\end{definition}

The correct way to handle $\tick$, if needed, in any of these models is to
add to the respective transformation in exactly the same way we did for stable
failures.  This is to be expected because $\tick$ only ever happens at the
end of traces.  Clearly we will need to use $term''$ as a cipher for
$\tick''$ in appropriate cases.

\subsection{Rational models}
We will later on wish to consider only models $\mmodel$ for which the correspondence
between $\flmodel$-observations and $\mmodel$ observations is decidable by a
finite memory computer.  We will interpret this notion as saying the the
relation $\mmodel_\Sigma$ corresponds to the language accepted by some finite
state automaton.  In order to do this, we must first decide how to convert
elements of $\flmodel_\Sigma$ to words in a language.  We do this in the obvious
way (the reasons for using fresh variables to represent the $A_i$ will become
apparent in Section \ref{sec:modshift}).

\begin{definition}The \emph{canonical encoding} of $\flmodel_\Sigma$ is over the
alphabet $\Xi :=\Sigma\cup\Sigma''\cup\Sym$, where
$\Sigma'':=\{a'':a\in\Sigma\}$ and
$\Sym=\{\langle,\rangle,`{\tt,}',\spot\}$.\footnote{Note that this
somewhat unsatisfactory notation denotes a set of four elements: the angle
brackets $\langle$ and $\rangle$, the comma , and the symbol $\spot$.}  It is
given by the representation in Definition \ref{def:flobs}, where sets $A_i$ are
expressed by listing the elements of $\Sigma''$ corresponding to the members of
$A_i$ in alphabetical order.  We denote this encoding by
$\phi_\Sigma:\flmodel_\Sigma\rightarrow\Xi^*$.\end{definition}

We now define a model to be \emph{rational} (borrowing a term from automata theory)
if its defining relation can be recognised (when suitably encoded) by some
nondeterministic finite automaton.

\begin{definition}\label{def:ratmod}A model $\mmodel$ is \emph{rational} if for every alphabet
$\Sigma$, there is some finite alphabet $\Theta$ and a map 
$\psi_\Sigma:\obs_\Sigma(\mmodel)\rightarrow\Theta^*$, such that there is a
(nondeterministic) finite automaton $\mathcal{A}$ recognising 
$\left\{\left(\phi_\Sigma(x),\psi_\Sigma(y)\right):(x,y)\in\mmodel_\Sigma\right\}$,
and such that $\psi_\Sigma$ is \emph{order-reflecting} (that is,
$\psi_\Sigma(x)\leq\psi_\Sigma(y)$ only if $x\leq y$), with respect to the
prefix partial order on $\Theta^*$, and the partial order induced by
$\mmodel_\Sigma$ on $\obs_\Sigma(\mmodel)$.
\end{definition}

What does it mean for an automaton to `recognise' a relation?

\begin{definition}\label{def:rec}For alphabets $\Sigma$ and $T$, a relation
$\mathcal{R}\subseteq\Sigma^*\times T^*$ is \emph{recognised} by an automaton
$\mathcal{A}$ just when:
\begin{enumerate}[(i)]
\item The event-set of $\mathcal{A}$ is $\lft.\Sigma\cup\rgt.T$, and
\item For any $s\in\Sigma^*, t\in T^*$, we have $s\mathcal{R}t$ if and only if
there is some interleaving of $\lft.s$ and $\rgt.t$ accepted by $\mathcal{A}$.
\end{enumerate}
\end{definition}

Note that recognisability in the sense of Definition \ref{def:rec} is easily
shown to be equivalent to the common notion of recognisability by a \emph{finite
state transducer} given for instance in \cite{shallit2009second}, but the above
definition is more convenient for our purposes.  Note also that $\flmodel$
itself (viewing $\flmodel_\Sigma$ as the diagonal relation) is trivially rational.

\begin{lemma}The models $\tmodel,\fmodel,\rmodel,\amodel$ and $\rtmodel$ are
rational.\end{lemma}

\noindent{\bf Proof}\\
By inspection of Definitions \ref{def:trmod}--\ref{def:rtmod}.  We take
$\Theta=\Sigma\cup\Sigma'\cup\Sigma''\cup\Sym$, with
$\Sigma''$ and the expression of acceptance sets as in the canonical encoding of
$\flmodel$, and refusal sets expressed in the corresponding way over
$\Sigma':=\{a':a\in\Sigma\}$.\et

Note that not all relations are rational.  For instance, the `counting relation'
mapping each finite linear observation to its length is clearly not rational.
We do not know whether the additional constraint of being a finite
observational model necessarily implies rationality; however, no irrational
models are known.  We therefore tentatively conjecture:
 that {\em every
finite observational model is rational.}

\section{Model shifting}\label{sec:modshift}
We now come to the main substance of this paper: we prove results on `model
shifting', showing that there exist contexts allowing us to pass between
different semantic models and the basic traces model.  The main result is 
Theorem \ref{thm:ratshift}, which shows that this is possible for any rational
model.

\subsection{Model shifting for $\mathcal{FL}$}
We begin by proving the result for the finest model, $\flmodel$.  We show that
there exists a context $\mathcal{C}_{\flmodel}$ such that for any process $P$,
the finite linear observations of $P$ correspond to the traces of
$\mathcal{C}_{\flmodel}(P)$.

\begin{lemma}[Model shifting for $\flmodel$]\label{lem:msfl}For every alphabet $\Sigma$, there
exists a context $\cfl$ over alphabet
$T:=\Sigma\cup\Sigma'\cup\Sigma''\cup\{\done\}$, and an order-reflecting map
$\pi:\flmodel_\Sigma\rightarrow T^*$ (with respect to the extension partial order
on $\flmodel_\Sigma$ and the prefix partial order on $T^*$) such that for any process $P$ over $\Sigma$
we have $\tmodel(\cfl[P]) = \pref(\pi(\flmodel(P)))$ (where
$\pref(X)$ is the prefix-closure of the set $X$).
\end{lemma}

\noindent{\bf Proof}\\
We will use the unprimed alphabet $\Sigma$ to denote communicated
events from the original trace, and the double-primed alphabet $\Sigma''$ to
denote (members of) stable acceptances.  $\Sigma'$ will be used in an intermediate step to
denote refusals, and $\done$ will be used to distinguish $\emptyset$
(representing an empty acceptance set) from $\spot$ (representing a failure to
observe anything).

\textbf{Step 1:} We first produce a process which is able to communicate events
$x_i'$, just when the original process can stably refuse
the corresponding $x_i$.  Define the partial order $\leq_1 = \langle x' <_1 x: x \in\Sigma\rangle$,
which prevents refusal events when the corresponding event can occur.

Let the context $\mathcal{C}_1$ be given by
\[\mathcal{C}_1[X] = \prior(X\interleave\RUN_{\Sigma'},\leq_1,\Sigma).\]
Note that the third argument prevents primed events from occurring in unstable
states.

\textbf{Step 2:} We now similarly introduce acceptance events, which can happen
in stable states when the corresponding refusal can't.  The crucial difference
between $a$ and $a''$ is that $a$ usually changes the underlying process state,
whereas $a''$ leaves it alone. $a''$ means that $P$ can perform $a$ from its
present stable state, but does not explore what happens when it does.

Similarly define the partial order $\leq_2 = \langle x'' <_2 x' :
x\in\Sigma\rangle$, which prevents acceptance events when the corresponding
refusal is possible.  Let the context $\mathcal{C}_2$ be defined by 
\[\mathcal{C}_2[X] =
\prior(\mathcal{C}_1[X]\interleave\RUN_{\Sigma''},\leq_2,\Sigma).\]

\textbf{Step 3:} We now ensure that an acceptance set inferred from a trace is
a complete set accepted by the process under examination. 
This is most straightforwardly done by employing a regulator process, which
can either accept an unprimed event or accept the alphabetically first refusal or
acceptance event, followed by a refusal or acceptance for each event in turn.  In the
latter case it then communicates a $\done$ event, and returns to its original
state.  It has thus recorded the complete set of events accepted by $P$'s
present state.

The $\done$ event is necessary in order to distinguish between a terminal $\emptyset$,
which can have a $\done$ after the last event, and a terminal $\spot$, which
cannot (observe that a $\emptyset$ cannot occur other than at the end).  Along the way, we hide the refusal events.

Let $a$ and $z$ denote the alphabetically (by which me mean in a fixed
but arbitrary linear order on $\Sigma$)
 first and last events respectively,
and let ${\tt succ }x$ denote the alphabetical successor of $x$.  Define the processes 
\[\begin{array}{rl}
	\unstp &= \Extchoice_{x\in\Sigma} x\then\unstp \\
&\qquad\extchoice a'\then\stabp(a) \extchoice a''\then\stabp(a) \\
\stabp(x) &= x'\then\stabp({\tt succ } x)\extchoice x''\then\stabp({\tt succ }
x)\qquad (x\neq z) \\
\stabp(z) &= \done\then\unstp,
\end{array}\]

and let

\[\cfl[X] =
\left(\mathcal{C}_2[X]\parallel[\Sigma\cup\Sigma'\cup\Sigma'']\unstp\right)\hide\Sigma'.\]

A
little care is required here.  We can prevent acceptances from being `skipped
over' by prioritising the double-primed events in alphabetical order, but we
also have to prevent acceptances from ending early, i.e. prevent an unprimed
event from happening prematurely.

The most obvious solution is to prioritise acceptance events over unprimed
events. This does not work, however, because the prioritise operator forces all
events which can be performed in unstable states to be maximal in the order,
and in the LTS representing the underlying process, any event
can happen as an alternative to $\tau$.

We instead use the event $\done$ to mark the end of an acceptance set, and use a
regulator process to prevent a double-primed event from being followed by an
unprimed event without an intervening $\done$.  This also distinguishes between
$\emptyset$, represented by a $\done$ between two unprimed events, and $\spot$,
represented by consecutive unprimed events.

Define the partial order 
\[\omega_3 = \langle x'' < y'' : y <_\alpha x\rangle \cup \langle \done < x'' :
x\in\Sigma\rangle,\]
which prevents jumps in acceptance sets, and allows $\done$ only in stable
states where no double-primed events are left to be communicated.  Let the
context $\mathcal{C}_3$ be defined by 
\[\mathcal{C}_3[X] =
\prior(\mathcal{C}_2[X]\interleave\RUN_\done,\omega_3,\Sigma).\]
We now define the regulator process which prevents sequences of double-primed
events not concluded by $\done$:
\[\begin{array}{rl}
	DREG &= \left( \Extchoice_{x\in\Sigma} x\then DREG\right)
\extchoice \left(\Extchoice_{x\in\Sigma} x''\then DREG'\right) \\
DREG' &= \left(\Extchoice_{x\in\Sigma}x''\then DREG'\right) \extchoice
\done\then DREG,
\end{array}\]
and then define the context $\cfl$ by
\[\cfl[X] = \mathcal{C}_3[X]
\parallel[\Sigma\cup\Sigma''\cup\{\done\}] DREG.\]

\textbf{Step 4:} We now complete the proof by defining the function $\pi$
inductively as follows:
\[\begin{array}{rl}
	\pi(s\cat\langle \spot\rangle ) &= \pi(s)\\
\pi(s\cat\langle x\rangle ) &= \pi(s)\cat\langle x\rangle \\
\pi(s\cat\langle A &=\{x_1,\ldots,x_k\}\rangle) = \pi(s)\cat\langle x_1'' \ldots
x_k''
\done\rangle,
\end{array}\]
where without loss of generality the $x_i$ are listed in alphabetical order.

It is clear that this is order-reflecting, and by the construction
above satisfies $\tmodel(\cfl[P]) =
\pref(\pi(\flmodel(P)))$.
\et

This result allows us to translate questions of $\flmodel$-refinement into
questions of trace refinement under $\mathcal{C}_{\flmodel}$, as follows:

\begin{corollary}\label{cor:flinc}For $\mathcal{C}_{\flmodel}$ as in Lemma \ref{lem:msfl}, and for
any processes $P$ and $Q$, we have $P\flrefinedby Q$ if and only if
$\cfl[P]\trefinedby\cfl[Q]$.
\end{corollary}

\noindent{\bf Proof}\\
Certainly if $\flmodel(Q)\subseteq\flmodel(P)$ then
$\tmodel(\cfl[Q]) = \pref(\pi(\flmodel(Q))) \subseteq \pref(\pi(\flmodel(P))) = 
\tmodel(\cfl[P])$ and so $\cfl[P]\trefinedby\cfl[Q]$.

Conversely, suppose there exists $x\in\flmodel(Q)\setminus\flmodel(P)$.  Then
since $\flmodel(P)$ is downwards-closed, we have $x\nleq y$ for all
$y\in\flmodel(P)$.  Since $\pi$ is order-reflecting, we have 
correspondingly $\pi(x)\nleq\pi(y)$ for all $y\in\flmodel(P)$.  Hence
$\pi(x)\notin\pref(\pi(\flmodel(P)))$, so $\pref(\pi(\flmodel(Q))) \nsubseteq
\pref(\pi(\flmodel(P)))$.
\et

\subsection{Model shifting for rational observational models}
We now have essentially all we need to prove the main theorem.  We formally
 record a
well known fact, that any Nondeterministic Finite Austomaton (NFA) can be implemented as a CSP process (up to
prefix-closure, since trace-sets are prefix-closed but regular languages
are not):

\begin{lemma}[Implementation for NFA]\label{lem:impl}Let
$\mathcal{A} = (\Sigma,Q,\delta,q_0,F)$ be a (nondeterministic) finite automaton.
Then there exists a CSP process $\PA$ such that
$\pref(L(\mathcal{A})) = \pref(\tmodel(\PA))$.
\end{lemma}
 See Chapter 7 of \cite{roscoe1997tpc} for the proof.

\begin{theorem}[Model shifting for rational models]\label{thm:ratshift}For every rational model $\mmodel$, there exists a context $\cm$
such that for any process $P$ we have $\tmodel(\cm[P])=\pref(\psi(\mmodel(P)))$.
\end{theorem}

\noindent{\bf Proof}\\
Let $\mathcal{A}$ be the automaton recognising $(\phi\times\psi)(\mmodel)$ (as
from Definition \ref{def:ratmod}), and let $\PA$ be the corresponding process
from Lemma \ref{lem:impl}.

We first apply Lemma \ref{lem:msfl} to produce a process whose traces correspond
to the finite linear observations of the original process, prefixed with $\lft$: let $\cfl$ be the
context from Lemma \ref{lem:msfl}, and let the context $\mathcal{C}_1$ be
defined by
\[\mathcal{C}_1[X] = \cfl[X]\rensubs{\lft.x}{x}.\]

We now compose in parallel with $\PA$, to produde a process whose traces
correspond to the $\mmodel$-observations of the original process.  Let
$\mathcal{C}_2$ be defined by
\[\mathcal{C}_2[X] = \left(\left(\mathcal{C}_1[X]\parallel[\{|\lft|\}]\PA\right)
\hide \{|\lft|\}\right)
\rensubs{x}{\rgt.x}.\]
Then the traces of $\mathcal{C}_2[X]$ are precisely the prefixes of the images
under $\psi$ of the observations corresponding to $X$, as required.
\et

By the same argument as for Corollary \ref{cor:flinc}, we have

\begin{corollary}\label{cor:ratref}For any rational model $\mmodel$, let $\cm$ be as in Theorem
$\ref{thm:ratshift}$.  Then for any processes $P$ and $Q$, we have
$P\refinedby[M]Q$ if and only if $\cm[P]\trefinedby\cm[Q]$.
\end{corollary}

\section{Implementation}
We demonstrate the technique by implementing contexts with the property of
Corollary \ref{cor:ratref}; source code may be
found at \cite{testsource}.

For the sake of efficiency we work directly rather than using the general
construction of Theorem \ref{thm:ratshift}.  The context {\tt C1} introduces
refusal events and a {\tt stab} event, which can occur only when the
corresponding normal events can be refused.  This implements the refusal testing
model, and the context {\tt CF} which allows only normal events optionally
followed by some refusals (and {\tt stab}) implements the stable failures model.

This is however suboptimal over large alphabets, in the typical situation where
most events are refused most of the time.  FDR4's inbuilt failures
refinement checking codes refusal in terms of \emph{minimal acceptance} sets (checking that each such
acceptance of the specification is a superset of one  of the implementation).
Minimal acceptances are typically smaller than maximal refusal sets.

For models based on acceptance sets rather than refusal sets, we have to
consider the whole collection of them rather than just the minimal ones.
We can introduce a second extra copy of $\Sigma$, namely $\Sigma''$, where
$a''$ will mean that the current stable state of $P$ can accept $a$, but the communication of $a''$ will not change the state of $P$.  We can do this essentially by applying the previous construction (creating $a'$) to itself.  This uses
an order $\leq''$ under which $a''<'a'$ on the process ${\cal C}(P)$:
\[{\cal C''}(P) = (\prior({\cal C}(P),\leq'',\Sigma\cup\Sigma')\parallel[\Sigma\cup\Sigma'\cup\Sigma''\cup\{stab\}]Reg''\]
Here $Reg''$ is a process that initially will accept a $\Sigma$ event or $a'$ or $a''$ for the alphabetically first member $a$ of $\Sigma$.  If either of
the latter it will insist on getting each subsequent member of $\Sigma$ in one
of these two forms until it has pieced together the complete acceptance set.
Thus as soon as the present state of $P$ is recognised as stable, $Reg''$
establishes its complete acceptance set before permitting to $P$ to carry on
further if desired. (For the acceptances model with only acceptances at
the ends of traces, there is no need to do so.)
As an alternative, $Reg''$ could communicate an event such as $done$ when it
gets to the end of the list of events, which would enable us to hide the
refusal events $\Sigma'$.

Similar constructions with slightly different restrictions on the permissible
sequences of events produce efficient processes for the revivals and refusal
testing models.  We will generalise this below.

\subsection{Testing}\label{sec:test}

We test this implementation by constructing processes which are first distinguished by
the stable failures, revivals, refusal testing and acceptance models respectively
(the latter two being also distinguished by the finite linear observations
model).  The processes, and the models which do and do not distinguish them, are
shown in Table \ref{tab:heir} (recall the precision hierarchy of models:
$\tmodel\leq\fmodel\leq\rmodel\leq\{\amodel,\rtmodel\}\leq\flmodel$).  The
correct results are obtained when these checks are run in FDR4 with the
implementation described above.
\begin{table}[htb]\small
\centering
\begin{tabular}{llll}
\hline
Specification & Implementation & Passes & Fails \\
\hline
$a\then\div$ & $a\then\STOP$ & $\tmodel$ & $\fmodel$ \\
$((a\then\div)\extchoice\div)\intchoice\STOP$ & $a\then\div$ & $\fmodel$ &
$\rmodel$ \\
$(a\then\div)\intchoice(\div\triangle(a\then\STOP))$ & $a\then\STOP$ & $\rmodel,\amodel$ &
$\rtmodel, \flmodel$ \\
$(a\then\STOP)\intchoice(b\then\STOP)$ & $(a\then\STOP)\extchoice(b\then\STOP)$
& $\rmodel,\rtmodel$ & $\amodel,\flmodel$ \\
\hline
\end{tabular}
\caption{Tests distinguishing levels of the model precision heir achy.
$\triangle$ is the \emph{interrupt} operator; see \cite{roscoe2010understanding}
for details.}
\label{tab:heir}
\end{table}
\subsection{Performance}
We assess the performance of our simulation by running those examples from Table
1 of \cite{gibson2015practical} which involve refinement checks (as opposed to
deadlock- or divergence-freedom assertions), and comparing the timings for our
construction against the time taken by FDR4's inbuilt failures refinement check
(since $\fmodel$ is the only model for which we have a point of comparison
between a direct implementation and the methods developed in this paper).  
Results are shown in Table \ref{tab:perf}, for both the original and revised contexts
described above; the performance of the $\flmodel$ check is also shown.  As may be seen, performance is somewhat worse but not
catastrophically so.  Note however that these processes involve rather small
alphabets; performance is expected to be worse for larger alphabets.
{\small
\begin{table}[htb]\small
\centering
\begin{tabular}{|l|ccc|ccc|ccc|ccc|} \hline 
&  \multicolumn{3}{|c|}{Inbuilt $\fmodel$} & \multicolumn{3}{|c|}{{\tt CF}} & 
\multicolumn{3}{|c|}{{\tt CF'}} & \multicolumn{3}{|c|}{{\tt FL}} \\
\hline
File & $|S|$ & $|\Delta |$ & $T(s)$ & $|S|$ & $|\Delta |$ &
$T(s)$ & $|S|$ &
$|\Delta |$ & $T(s)$ & $|S|$ & $|\Delta |$ & $T(s)$\\ 
\hline 
{\tt inv} &  21 & 220 & 23 & 21 & 220 & 78 & 21 & 220 & 125 & 21 & 220 &
145\\
{\tt nspk}  & 6.9 & 121 & 22 & 6.3 & 114 & 73 & 4.1 & 72 & 55 & 5.4 &
97M & 92\\
{\tt swp} & 24 & 57 & 16 & 30 & 123 & 61 & 43 & 76 & 107 & 42 & 93 & 131 \\ \hline
\end{tabular}
\caption{Experimental results comparing the performance of our construction with
FDR3's inbuilt failures refinement check.  $|S|$ is the number of states,
$|\Delta|$ is the number of transitions, $T$ is the time (in seconds), 
all state and transition counts are in millions.}
\label{tab:perf}
\end{table}
	}
\subsection{Example: Conflict detection}
We now illustrate the usefulness of richer semantic models than just traces and
stable failures by giving a sample application of the revivals model.  Suppose that we have a
process $P$ consisting of the parallel composition of two sub-processes $Q$ and
$R$.  The stable failures model is able to detect when $P$ can refuse all the events of
their shared alphabet, or deadlock in the case when they are synchronised on the whole
alphabet.  
However, it is unable to distinguish between the two possible causes
of this: it may be that one of the arguments is able to refuse the entire
shared alphabet, or it may be that each accepts some events from the shared
alphabet, but the acceptances of $Q$ and $R$ are disjoint.  We refer to the
latter situation as a `conflict'.  The absence of conflict (and similar
situations) is at the core of a number of useful ways of proving
deadlock-freedom for networks of processes running in
parallel~\cite{deadlockfreedom1988}.

The revivals model can be used to detect conflicts.
For a process $P=Q\parallel[X][Y]R$, we introduce a fresh event $a$ to
represent a generic event from the shared alphabet, and form the process
$P'=Q'\parallel[X'][Y']R'$, where $Q'=Q\rename{\{(x,x),(x,a):x\in X\}}$,
$X'=X\cup\{a\}$,
and similarly for $R'$ and $Y'$.  Conflicts of $P$ now correspond to revivals
$(s,X\cap Y,a)$, where $s$ is a trace not containing $a$.

\section{Timed Failures and Timed CSP}
Timed CSP is a notation which  adds a $WAIT\;t$ construct  to CSP and reinterprets
how processes behave in a timed context.  So not only does it constrain the order that things happen, but also when they happen.  Introduced in~\cite{rr}, it
has been widely used and studied~\cite{sch,schop,davies1995brief}. $WAIT\:t$ behaves like $SKIP$
except that termination takes place exactly $t$ time units after it starts.    It introduced and uses
the vital principle of {\em maximal progress}, namely that no action that
is not waiting for some other party's agreement is delayed: such actions do
not sit waiting while time passes.  That principle fundamentally changes 
the nature of its semantic models.  

Consider how the hiding operator is defined.
It is perfectly legitimate to have a process $P$
 that offers the initial visible
events $a$ and $b$ for an indefinite length of time, say $P=a\then P1\extchoice b\then P2$.  However $P\hide \{a\}$
cannot perform the initial
 $b$ at any time other than the very beginning because the
$a$ has become a $\tau$.  So $P\hide X$ only uses those behaviours of $P$ which
refuse $X$ whenever time is passing.

Timed CSP was originally described on the basis of continuous (non-negative real) time values.  The basic unit of semantic discourse is a {\em timed failure},
the coupling of a timed trace~-- a sequence of events with non-strictly increasing times~-- and a timed refusal, which is the union of a suitably finitary
products of a half-open time interval $[t_1,t_2)$ (containing $t_1$ but not
$t_2$) and a set of events.  Thus the refusal set changes only
finitely often in a finite time, coinciding with the fact that a process can
only perform finitely many actions in this time.    This continuous
model of time takes it well outside the finitary world that model checking finds comfortable.  However it has long been known that restricting the $t$
in $WAIT\;t$ statements to integers makes it susceptible to a much more
finitary analysis by region graphs~\cite{jackson}.  However the latter
represents a technique remote from the core algorithms of FDR so it has never
been implemented for CSP, though it has for other notations~\cite{up}.  In~\cite{ou1,ou2}, Joel
Ouaknine made the following  important discoveries:
\begin{itemize}
\item  It makes sense to interpret Timed CSP with integer $WAIT$ over the
positive integers as time domain.
\item  The technique of {\em digitisation} (effectively a uniform mapping
of general times to integers) provides a natural mapping between these
two representations. 
\item  Properties that are {\em closed under inverse digitisation} can be decided
over continuous Timed CSP by analysis over Discrete Timed CSP, and these
include many practically important specifications.
\item  It is in principle
 possible to interpret Discrete Timed CSP in a modified  (by the addition of
two new operators)  $tock$-CSP (a dialect
		developed by Roscoe in the early 1990's for reasoning about timed systems in FDR)  and therefore  in  principle it is possible to reason about continuous Timed CSP in FDR.  The definition of Timed CSP hiding over LTSs
		involves prioritising $\tau$ and $\tick$ over $tock$.

\end{itemize}

This was implemented as described in
in~\cite{mctcsp}, originally in the context of the last
versions of FDR2 and Timed CSP
 continues to be supported in FDR4.  There is an important thing missing from these implementations, however, namely refinement checking in the Timed Failures Model, the details of which we describe below.  That means that although it
is possible to check properties of complete Timed CSP systems, there is no
satisfactory compositional theory for (Discrete) Timed CSP.   For example one
cannot automate the reasoning that if $C[P,Q]$ (a term in Timed CSP) satisfies $SPEC$, and
$P\sqsubseteq P'$ and $Q\sqsubseteq Q'$ then $C[P',Q']$ satisfied $SPEC$, because FDR does not give us a means of checking the necessary refinements.

The purpose of this section is to show how Timed Failures refinement can be
reduced to things FDR can do, filling this hole.  Given the methods
described in this paper to date, it is natural to try model shifting, and
we will do this below.  There is another option offered to us by late
versions of FDR2, namely reduction to the Refusal Testing model which is
implemented in that but not (at the time of writing) later versions of FDR.
We will discuss these in turn.
\subsection{A summary of Discrete Timed Failures}
The Discrete Timed Failures model $\cal D$  consists, in one presentation, of
sequences of the form
\[(s_0,X_0,tock,s_1,X_1,tock,\ldots,s_{n-1},X_{n-1},tock,s_n,X_n)\]
where each of $s_i$ is a member of $\Sigma^*$, each of $X_i$ is  a subset of $\Sigma$, and $tock\not\in\Sigma$.  Since $tock$ never happens from an
unstable state, there is no need to have the possibility  of $\bullet$ as
discussed above  for  other models before $tock$, and it would be misleading to have it.  
We do however allow $\bullet$ for $X_n$.

What this means, of course, is that the trace $s_0$ occurs, after which it
reaches a stable state where $tock$ occurs, and this is repeated for other
$s_i$ and $X_i$ until, after the last $tock$,  the trace  $s_n$ is performed
followed by the refusal $x_n$ (not including $tock$) or potentially instability.
Recall that we apply the principle of maximal progress, so that $tock$ only
happens from a stable state: this means that if, after behaviour $\ldots s_n$, stability is not observable, then $tock$ can never happen and we have reached an
error state.  It is, however, convenient to have this type of error state in our
model because misconstrued systems can behave  like  this.

Like other CSP models, it has healthiness conditions, or in other words
properties that the representation of any real process must satisfy.  These
are analogous to those of related untimed models, such as prefix closure and
subset closure on refusal sets, and the certain refusal of impossible
events.  A property that it inherits from continuous Timed CSP is {\em no
instantaneous withdrawal}, meaning that if, following behaviour $\beta$, it
is impossible for a process to refuse $a$ leading up to the next $tock$, then
the process must still have the possibility of performing $a$ after $\beta\trace{tock}$.  This amounts to the statement that the passage of time as
represented by $tock$ is not directly visible to the processes concerned,
and is much discussed in the continuous context in~\cite{tsm,tni}.

$\cal{D}$ is a rational model, since it can be obtained from the standard 
representation of $\rtmodel$ by the rational transduction which deletes all 
refusal sets preceding events other than $tock$ (and replaces non-terminal 
occurences of $\spot$ by $\emptyset$, since $tock$ can only occur in stable states).
Hence by Theorem \ref{thm:ratshift} it can be model shifted: there exists a 
context $\mathcal{C}_{\mathcal{D}}$ such that trace refinement under 
$\mathcal{C}_{\mathcal{D}}$ is equivalent to refinement in $\cal{D}$.

The operational semantics of Discrete Timed CSP processes, under the
transformation described and implemented in~\cite{mctcsp}. Have the property that
$tock$ is available in every stable state and no unstable state.

\subsection{Model shifting Timed Failures}

We can capture this through model shifting by introducing a primed  copy $a'$ of 
each $a\in\Sigma$ and using the  following construct involving
a  regulator which ensures
that an ordinary event cannot follow a refusal flag.  This means
that its traces consist of pairs of traces of $\Sigma$ and traces of
$\Sigma'$ (which can be empty) interspersed with a $tock$ between
consecutive pairs.
\[\begin{array}{rcl}
CS_{TF}(P) &=& \prior(\leq,P\interleave  RUN(\Sigma'),\Sigma)\parallel Reg\\[2ex]
Reg&=& tock\then Reg\\
	&&\extchoice (\Extchoice_{a\in\Sigma} a\then Reg)\\
	&&\extchoice (\extchoice_{a\in \Sigma} a'\then Reg1)\\[1ex]
Reg1&=& tock\then Reg\\
	&&\extchoice (\Extchoice_{a\in \Sigma} a'\then Reg1)
\end{array}\]
and $a'<a$ (as well as the implicit $a'<\tau$) for each $a\in\Sigma$.  

We have assumed here that any prioritisation needed to ensure maximal
progress has already been  applied before  this, so that the LTS
being operated on here has the correct behaviour under a normal
interpretation.

Note that this  regulator  allows only allows refusal  events and $tock$
after refusal events $a'$, thus forcing  the decorated traces (namely
combinations  of  real events and the $a'$ ones signifying refusals)  to
exactly follow the structure set out for timed  failures above.  Thus aside
from  the exact structure of the model, we  have followed the same procedure as that
used for the stable  failures  model  above.

Note  that
\begin{itemize}
\item  Events  in $\Sigma$  cannot follow refusal (primed events); only  other
primed events or $tock$.
\item  There  would  have been no harm in using the $stab$ event seen for
	stable failures (perhaps most elegantly so that it can only
		happen as the last event),  but for Timed CSP processes  this would make no difference to the equivalence
or refinement relations induced.  This is because $stab$ would be possible
after a trace if and only if $tock$ is.
\item Timed Failures refinement between  two Timed CSP processes is decided  by
traces refinement between the decorated and  regulated  transformed processes
as defined above.  Thus  it does not matter (when using them for this
purpose) that the regulator adds in further refusals not possible for the
original process (namely, after any $a'$ event, the regulated process
refuses the whole of $\Sigma$.
\end{itemize}

\subsection{Reducing Timed Failures to refusal testing}
In effect the Timed Failures model is the refusal testing model with all
refusal sets that precede a  non-$tock$ event ignored.  If follows that if
we can create a context $C[\cdot]$  in CSP such that  $C[P]$  contains 
precisely the behaviours of $P$ that should not be forgotten, then we can
state that $C[P]\sqsubseteq_{RT}  C[Q]$ if and only if $P\sqsubseteq_{TF}Q$.

This  seems  difficult, not  least because when observing $P$ refusing something, we cannot stop it performing any action other than $tock$ without affecting
the refusal itself.  We can again solve this by use of a regulator process which
allows any $\Sigma$ event to happen at any time from an unstable state and carry
on, or from a stable state allow any $\Sigma$ event leading to the divergent
process $DIV$, or $tock$ after which the regulator just carries on.
\[\begin{array}{rcl}
RE GP&=&(\Extchoice_{a\in \Sigma} a\then RE GP)\\
&&\rhd\\
&&(tock \then RE GP\\
&&\extchoice \Extchoice_{a\in\Sigma} a\then DIV)
\end{array}\]
$RE GP$ is thus a three-state process: in the initial state all the $\Sigma$ 
events can happen as can a $\tau$  taking it to the second and stable state.
The third state is $DIV$.  Note how unstable states and the fact that
$DIV$  is refinement-maximal in the model are crucial in making this construction work.  As before, this regulator is synchronised with $P$ to perform the 
transformation.

So we have defined a projection
\[\Pi_{TF}(P)  = P\parallel  RE GP\]

This and $MS TF(P)$ are thus faithful representations of the timed failures semantics of $P$  in two different models.  They can be used for comparisons under
refinement in these models.

Because $\Pi_{TF}$ simply records behaviours with refusals before members of $\Sigma$
from all processes,  we notice that in general
\[P\sqsubseteq_{TF}  Q\Leftrightarrow  \Pi_{TF}(P)\sqsubseteq_{RT}\Pi_{TF}(Q)
\Leftrightarrow  P\sqsubseteq_{RT}\Pi_{TF}(Q)\]

\section{Case study: Timed Sliding Window Protocol}
The sliding window protocol has long been used as a case study with FDR: it
is well known and reasonably easy to understand, at least in an untimed setting.
It is a development of the alternating bit protocol in which  the messages in a fixed-length  window on the
input stream are simultaneously available for transmission and acknowledgement
across an erroneous medium which, in our version, can lose and duplicate
messages but not re-order them.  We have re-interpreted this in Timed CSP with
the following features:
\begin{itemize}
\item There is a parameter $W$ which defines the width of the window.  Because
the windows held by the sender and receiver processes may be out of
step, we need to define $B=2W$ to be the bound on the
amount of buffering the system can provide.
\item In common with other CSP codings of this protocol, we need to
make the indexing space of places in the input and output streams finite
by replacing the natural non-negative integers by integers modulo some
		$N$ which must be at least $2W$ (though there is no requirement that $B$ and $N$ are the same).   This is sufficient to ensure that acknowledgement tags never get confused as referring to the wrong message.
\item Round robin sending of message components from unacknowledged items in the
	current window: this clearly has a bearing on the timing behaviour of
the transmission and acknowledgements that the system exhibits.
\item The occurrence of errors is limited by a parameter which forces them to
	be spaced: at least $K$ time units must pass between consecutive ones.
		To achieve this elegantly we have used the controlled error model~\cite{roscoe1997tpc} in which errors are triggered by events that can be restricted by
		external regulators, and then lazily abstracted.  It turns out
that lazy abstraction (originally proposed in~\cite{roscoe1997tpc}) needs reformulating
in Timed CSP.  We will detail this below.

Clearly it would be possible to use different error assumptions.
\item We have assumed for simplicity that all ordinary actions take one
	time unit to complete.
\item Where a message is duplicated, we need to assume that the duplicate is
	available reasonably quickly, say within 2 time units of the original
		send.  If it can be deferred indefinitely this causes subtle
		errors in the sense that deferred duplication can prevent the
		system from settling sufficiently.
\end{itemize}

We can create a Timed Failures specification in CSP  which says, following established
models for regular CSP, that the resulting system is a buffer bounded by
$B$ (so it never contains more than $B$ items) but is only obliged
to input when it has nothing in it.  Whenever it is nonempty it is obliged
to output, but these two obligations do not kick in before some parameter $D$
time units from the previous external communication.  

This is slightly trickier than we might think because of the way in which
the implementation process can entirely legitimately change its behaviour over time.  So in an interval where it can legitimately accept or refuse an input
$left.1$, at one point it can refuse to  communicate it, while later accepting
it after time has passed.  

In hand-coded $tock$-CSP this can be expressed as
\begin{verbatim}
TFBUFF(n) =
let
TFB(s,k) =
 if k < n then
   ((#s>0 & right!head(s) -> TFB(tail(s),0)
              []
    #s<B & left?x -> TFB(s^<x>,0))
    [>
    tock  -> TFB(s,k+1))
  else
    ((#s>0 & right!head(s) -> TFB(tail(s),0))
    []
    (#s==0 & left?x -> TFB(<x>,0))
    []
    ((#s>0 and #s<B) & (left?x -> TFB(s^<x>,0) [> STOP))
    [] tock -> TFB(s,k))
    within TFB(<>,0)
\end{verbatim}
This says that if we have not yet reached the point where offers must be
made (i.e. \verb+k < n+) then it can perform permitted actions but can
(expressed via \verb+[>+ or sliding choice) also refuse them and
wait for time to pass.

In Timed CSP a completely equivalent specification can be divided into
three separate parts: one to control the buffer behaviour, one to 
handle what the specification says about when offers must be made as opposed
to can be made, and the final one to control nondeterminism by creating
the  most nondeterministic timed process on a given alphabet.  The last of these
is  notably trickier than in the untimed world because where a process
has the choice,  over a period, to accept or refuse an event $b$, it is not
sufficient for it to make the choice once and for all.  So we have 
\begin{verbatim}
TCHAOS(A) = let onestep = ([] x:A @ x -> onestep)
                          [> WAIT(1)
            within onestep;TCHAOS(A)
	    \end{verbatim}
In Timed CSP lazy abstraction needs to be formulated with this revised
$Chaos$ definition
\begin{verbatim}
LAbs(A)(P) = (P [|A|] TCHAOS(A))\A
\end{verbatim}
noting that the passage of time ($tock$) is implicitly synchonised here
as well as $A$, and priority of $\tau$ over $tock$ will also apply.

	    In the main part of the buffer specification we do not create this
	    style of nondeterminism, but instead use two variants of the externally visible events: one that will be made nondeterministic by the above
	    and one that will not:
\begin{verbatim}
TFB(s) =
   (#s>0 & right!head(s) -> TFB(tail(s)))
    []
   (#s>0 & rightnd!head(s) -> TFB(tail(s)))
    []
    (#s==0 & left?x -> TFB(<x>))
    []
    (#s<B & (leftnd?x -> TFB(s^<x>)))
\end{verbatim}
The above always allows the nondeterministic variants of the events, and
allows the ''deterministic'' ones when they should be offered if sufficient time
has passed since the last visible event.  Thus \verb+left+ is only
offered deterministically when the buffer is empty, no matter how long
since the last event.

The choice over whether the offers available must be made, implemented by
allowing the deterministic versions of events, is made by the following process
\begin{verbatim}
TEnable(E,R,m) = 
let Rest = diff(R,E)
    En = [] x:R @ x -> Dis(m)
 Dis(k) = if k==0 then En else
         (([] x:Rest @ x -> Dis(m))
         [] WAIT(1);Dis(k-1))
within Dis(m)
\end{verbatim}
The three parameters here are the events that are enabled when there has been
sufficient delay (here \verb+{|left,right|}+), the ones that reset the
clock (here \verb+{|left,right,leftng,rightnd|}+ and the time by
which offers have to be made.  The full specifiation is put together by
combining the above process, \verb+TFB(<>)+ and \verb+TCHAOS({|leftnd,rightnd|})+ and renaming \verb+leftnd, rightnd+ to respectively \verb+left, right+.

Given the subtlety of the above and the fact that it is hard to be
sure that \verb+TBUFF+ is right when it is written in $tock$ CSP rather than
Timed CSP, it is reassuring that FDR readily proves that the two versions
of the specification are equivalent in the Timed Failures model.

The authors have run a number of checks of versions of the Timed CSP version
of the protocol against this and other specifications.  
When compared against FDR's inbuilt stable failures refinement (a less
discerning one than timed failures, so not always producing the same
results) the overheads were low, typically about 50\% states and time.

\subsection{Experiments}
Files illustrating this section can be downloaded\footnote{\tt http://www.cs.ox.ac.uk/people/publications/personal/Bill.Roscoe.html}.

The following reports on the check of the Timed CSP sliding window protocol with
two items of $DATA$ and a window of width $4$ against the specification
that says it is an $8$-bounded buffer when there is
a minimum time between errors
of $3$.  It is specified to make stable offers by 42 time units. (In general
the longer between errors, the faster the system makes settled offers.)

  The first
check does this by model shifting, but it fails when the check is nearly complete because it can fail to have the offer ready on time.  In fact the
corresponding check is passed when 42 is replaced by 45 (but not 44).
\begin{verbatim}
assert CTFMS(TFBUFF(42)) [T= CTFMS(TLAbs({loss,dup})(ELSYSTEM(3)))
\end{verbatim}

The statistics from this check were as follows:

{\em Visited 49,239,989 states and 166,698,488 transitions in 118.91 seconds (on ply 261}

The following is a failures check of the same system without model shifting,
which happens to find the same problem.

\begin{verbatim}
assert TFBUFF(42) [F= TLAbs({loss,dup})(ELSYSTEM(3))
\end{verbatim}

{\em Visited 41,779,778 states and 107,648,549 transitions in 81.64 seconds (on ply 261)}

The following are the statistics from the same check simplified to a no-model shifting traces
check, which does not find the problem, and so passes.

{\em Visited 15,413,107 states and 36,428,632 transitions in 19.71 seconds (on ply 186)}

The smaller state count here is probably mainly
 because the normalised specification
in this final case is significantly smaller, as the count-down to forcing
an offer is irrelevant to traces.

It is noteworthy that the overhead of  model shifting (relatively speaking)
is here less than reported earlier for the untimed case.  We expect this is explained
because the unshifted checks in the timed case already contain (timed)
prioritisation
before it is applied as part of model shifting.

The experiments in this section were performed on a MacBook with a 2.7GHz Intel Core i7 processor.

\section{Conclusions}
We have  seen how the expressive power of CSP, particularly when extended
by priority, allows seemingly any finite behaviour model of CSP to be
reduced  to traces.  Indeed this extends to any finitely expressed rules
for what can be observed within  finite linear behaviours, whether the
resulting equivalence is compositional or not.  

This considerably extends the range of what can be done with a tool like
FDR.  The final section shows an alternative approach to this, namely
reducing a less discerning model to a more discerning one without priority.
This worked well for reducing timed failures to refusal testing, but other
reductions (for example ones involving both acceptances and refusal sets)
do not always seem to be so efficient.  For example reducing a refusal
sets process to the acceptances model seems unnecessarily complex as, for
example, the process $CHAOS$ needs exponentially many acceptance sets where
a single maximal refusal suffices.

We discovered that it is entirely practical to use this technique 
to reason about large systems.  Furthermore the authors have found
that the debugging feedback that FDR gives to model shifting checks
is very understandable and usable.  

In particular the authors were pleased to find that the results
of this paper make automated reasoning about Timed CSP practical.  They have
already found it most informative about the expressive power of the
notation.  It seems possible that, as with untimed CSP, the availability of
automated refinement checking will bring about enrichments in the notations
of Timed CSP that help it in expressing practical systems and specifications.

Model shifting means that it is far easier to experiment with automated
verification in a variety of semantic models, so it will only 
very occasionally be necessary for a new one to be directly supperted.

We believe that similar considerations will apply to classes of
models that include infinite observations such as divergences, infinite
traces, where these can be extended to incorporate refusals and acceptances
as part of such observations.  In such cases we imagine that model shifting
will take care of the aspects of infinite behaviours that
are present in their finite prefixes, and that the ways that
infinitary aspects are handled will follow one of the three traces
models available in CSP.  These are
\begin{itemize}
	\item finite traces (used in the present paper), 
	\item divergence-strict finite and infinite traces, so as soon as an observation is made
that can be followed by immediate divergence, we deem all continuations
to be in the process model whether or not the process itself can
do them operationally, and finally
		\item with full divergence strictness replaced by
the weak divergence strictness discussed in~\cite{ds} (here an infinite
behaviour with infinitely many divergent prefixes is added as above).
\end{itemize}

Thus it should be possible to handle virtually the
entire hierarchy of models described in~\cite{roscoe2010understanding} in terms of variants on
traces and model shifting.  This will be the subject of future research.

\subsubsection*{Acknowledgements}
The authors are grateful to Tom Gibson-Robinson for helpful discussions and
practical assistance with FDR4.  This work has been partially sponsored by DARPA
under agreement number FA8750-12-2-0247 and by a grant from EPSRC.

\bibliography{mybib}{}
\bibliographystyle{plain}
%
%\newpage
%\section*{Appendix (for review only)}
%The following is the source code used for the correctness tests described in
%Section \ref{sec:test}:
%\lstinputlisting[basicstyle=\scriptsize\ttfamily]{testheirarchy.csp}
\end{document}